\documentclass[showpacs,amsmath,amssymb,prd,nofootinbib,twocolumn]{revtex4}

\usepackage{graphicx}
\usepackage{dcolumn}
\usepackage{bm}
\usepackage{epsf}
\usepackage{hyperref}
\usepackage{amsfonts}
\usepackage{verbatim}
 
\begin{document}
 
\title{Approximate initial data for binary black holes}

\author{Kenneth A.~Dennison}

\affiliation{Department of Physics and Astronomy, Bowdoin College,
Brunswick, Maine 04011}

\author{Thomas W. Baumgarte}

\altaffiliation{Department of Physics, University of Illinois at
        Urbana-Champaign, Urbana, IL, 61801}

\affiliation{Department of Physics and Astronomy, Bowdoin College,
Brunswick, Maine 04011}

\author{Harald P. Pfeiffer}

\affiliation{Theoretical Astrophysics, California Institute of
Technology, Pasadena, California 91125}

\date{{\rm draft of} \today}

\begin{abstract}
We construct approximate analytical solutions to the constraint
equations of general relativity for binary black holes of arbitrary
mass ratio in quasicircular orbit.  We adopt the puncture method to
solve the constraint equations in the transverse-traceless
decomposition and consider perturbations of Schwarzschild black holes
caused by boosts and the presence of a binary companion.  A
superposition of these two perturbations then yields approximate, but
fully analytic binary black hole initial data that are accurate to
first order in the inverse of the binary separation and the square of
the black holes' momenta.  
\end{abstract}

 
\maketitle

\section{Introduction}
\label{INTRO}

Binary black holes are among the most promising sources of
gravitational radiation for the Laser Interferometer Gravitational
wave Observatory (LIGO) and other gravitational wave interferometers.
LIGO has recently reached its design sensitivity, making
the detection of astrophysical sources of gravitational radiation a
distinct possibility.  To aid both in the identification of such a
signal and in its interpretation, theoretical templates of potential
gravitational wave forms are urgently needed.

Numerical relativity is the most promising tool for modeling the
coalescence and merger of binary black holes.  Typically, numerical
relativity calculations compute a solution to Einstein's equations in
two steps.  In the first step, the constraint equations of general
relativity are solved to construct initial data, describing a snapshot
of the gravitational fields at one instant of time, and in the second
step these initial data are evolved forward in time by solving the
evolution equations.

Different approaches have been used to construct binary black hole
initial data (see, e.g., the reviews \cite{Coo00,BauS03} and
references therein).  Most earlier calculations employed the so-called
transverse-traceless decomposition of the constraint equations
(e.g.~\cite{Coo94,Bau00,PfeTC00,TicB04}, see also below), while most
more recent calculations solve the constraint equations in the
conformal thin-sandwich formalism
(e.g.~\cite{GouGB02,GraGB02,CooP04,CauCGP06}; see also
\cite{Alv00,YunTOB05,YunT06, Nis06} for alternative ways of
constructing binary black hole initial data.)  There is general
consensus that the latter formalism is better suited for the
construction of quasiequilibrium data (but see \cite{ShiUF04} for a
very promising alternative approach), even though, at least in terms
of global quantities, both formalisms lead to very similar results for
configurations outside the innermost stable circular orbit (see, e.g.,
\cite{TicBL03,CooP04}).

Dynamical evolutions of binary black holes have long suffered from
numerical instabilities.  The past year, however, has seen dramatic
progress, and at this point independent codes using very different
approaches and techniques can reliably model the coalescence, merger
and ring-down of binary black holes
\cite{Pre05b,CamLMZ06,BakCCKM06,DieHPSSTTV06,HerSL06}.  Some of these
calculations \cite{CamLMZ06,BakCCKM06,HerSL06} treat the black hole
singularity with the so-called puncture method \cite{BraB97} (see also
\cite{BeiO94,BeiO96}).  Given that puncture initial data for binary
black holes are most easily constructed in the transverse-traceless
formalism (see \cite{Bau00} and compare \cite{HanECB03,Han05}), this
development has renewed some interest in transverse-traceless initial
data.

In this paper we construct black hole binary initial data for
arbitrary mass ratios perturbatively.  We adopt the
transverse-traceless decomposition together with the puncture approach
and treat both the effect of each black hole's boost as well the
effect of the companion as a perturbation of a spherically symmetric
Schwarzschild black hole.  Both effects are axisymmetric --albeit with
different axes of symmetry-- allowing us to find simple analytic
expressions for the gravitational fields, the location of the apparent
horizon, the irreducible mass, and the total energy.  To leading order
we can construct binaries by simply adding the individual corrections,
resulting in analytic, perturbative black hole initial data.  The
solution is accurate up to order $(P/M)^2$ and $(M/s)$, where $P$ is
the black hole's momentum, $M$ a measure of the black hole's mass, and
$s$ the binary separation, and becomes exact in the limit of infinite
separation.  We find Newtonian expressions for the energy.  While this
is not surprising, it is certainly reassuring.  Recovering the
Newtonian limit is also non-trivial, since it requires taking into
account the effect of both the boost of the black holes, and the
distortion due to the companion black hole.

The numerical construction of initial data for binary black holes in
quasi-circular orbit requires significant computational resources,
since it involves stepping through nested iterations of elliptic
solves, root-finding and sometimes even minimizations along certain
sequences (see \cite{Coo94,Bau00,CauCGP06} and Section \ref{QCO}
below).  Our perturbative framework allows for a completely analytical
treatment, making the construction of approximate binary black hole
initial data for arbitrary mass ratios remarkably simple.  Our results
may therefore be of interest as initial data for dynamical
simulations, especially for large binary separations where the errors
are relatively small.  Perhaps more importantly our results provide
analytical insight into the structure of black hole binaries.  Several
of our intermediate results have been derived previously
(e.g.~\cite{CooY90,Bei00,Lag04, SchKB06}), but to the best of our
knowledge they have never been combined to construct binary black hole
initial data.

Our paper is organized as follows.  In Section \ref{IVP} we briefly
review the initial value problem of general relativity and the
transverse-traceless decomposition of the constraint equations in
particular.  We present perturbative solutions of boosted black holes,
of black holes with companions, and finally of binaries in
quasicircular orbit in Section \ref{PERT}.  In Section \ref{cookbook}
we outline how the results of the previous Sections can be used to
construct approximate binary black hole initial data for black holes
of arbitrary mass ratio in quasicircular orbit.  We conclude with a
discussion and summary in Section \ref{SUM}.  We also provide several
Appendices that contain all derivations and details omitted in the
main body of the text.  Throughout this paper we adopt geometrical
units in which $G = c = 1$.


\section{The Initial Value Problem}
\label{IVP}

Under a 3+1 decomposition~\cite{ArnDM62,Yor79}, Einstein's equations
of general relativity split into a set of constraint equations --the
Hamiltonian constraint and the Momentum constraint-- and a set of
evolution equations.  Constructing a set of initial data requires
specifying a spatial metric $\gamma_{ij}$ and an extrinsic curvature
$K_{ij}$ on a spatial hypersurface $\Sigma$ that satisfy the
constraint equations.  Such solutions are usually constructed with the
conformal method, whose most general form is given in~\cite{Yor99,
PfeY03}.  Here, we investigate a special case of the general
formalism, which was known earlier, and which is amenable to
analytical treatment.  We introduce a conformal transformation of the
spatial metric
\begin{equation}
\gamma_{ij} = \psi^{4} \hat \gamma_{ij},
\end{equation}
where $\psi$ is the conformal factor and $\hat \gamma_{ij}$ the
conformally related background metric.  We also split the extrinsic
curvature $K_{ij}$ into its trace $K$ and a conformally rescaled
trace-free part $\hat A_{ij}$ according to
\begin{equation}
K_{ij} = \psi^{-2}\hat A_{ij} + \frac{1}{3}\gamma_{ij}K.
\end{equation}
In terms of these variables the Hamiltonian constraint becomes
\begin{equation}
8\hat \nabla^{2} \psi-\psi\hat R-\frac{2}{3}\psi^{5}K^{2} + 
\psi^{-7}\hat A_{ij}\hat A^{ij} = 0,
\end{equation}
and the momentum constraint
\begin{equation}
\hat D_{j}\hat A^{ij} - \frac{2}{3}\psi^{6}\hat \gamma^{ij}\hat D_{j}K = 0.
\end{equation}
Here $\hat \nabla^{2} = \hat \gamma^{ij} \hat D_i \hat D_j$ is the
Laplacian, $\hat D_{i}$ the covariant derivative, and $\hat R$ the
Ricci scalar associated with the background metric $\hat \gamma_{ij}$,
and we have also assumed vaccuum.  

Both the conformal background metric $\hat \gamma_{ij}$ and the trace
of the extrinsic curvature $K$ remain freely specifiable.  We choose
conformal flatness $\hat \gamma_{ij} = f_{ij}$, where $f_{ij}$ is the
flat metric, and maximal slicing $K=0$, in which case the constraint
equations reduce to
\begin{equation}
\label{hconsimple}
\hat \nabla^{2}\psi = -\frac{1}{8}\psi^{-7}\hat A_{ij}\hat A^{ij}
\end{equation}
and
\begin{equation}
\label{pconsimple}
\hat D_{j}\hat A^{ij} = 0.
\end{equation}
Quite remarkably, the momentum constraint (\ref{pconsimple}) becomes
linear and decouples from the Hamiltonian constraint
(\ref{hconsimple}).  An analytical ``Bowen-York'' solution to the
momentum constraint, describing a black hole at coordinate location
${\bf C}$ with linear momentum ${\bf P}$ is given by \cite{Bow79,BowY80,Yor89}
\begin{equation} \label{BYAij}
\hat A^{ij}_{{\bf CP}}=
\frac{3}{2r_{{\bf C}}^2}\left[P^{i}n^{j}_{{\bf C}}+
P^{j}n^{i}_{{\bf C}}-\left(f^{ij}-n^{i}_{{\bf C}}n^{j}_{{\bf C}}\right)
P_{k}n^{k}_{{\bf C}}\right],
\end{equation}
where $r_{{\bf C}}=\left\|x^{i}-C^{i}\right\|$ is the coordinate
distance from the center of the black hole and $n^{i}_{{\bf
C}}=\left(x^{i}-C^{i}\right)/r_{{\bf C}}$ is the unit vector pointing
from that center to coordinate location ${\bf x}$.  Given the 
linearity of the momentum constraint we can construct solutions for
two boosted black holes by simple superposition,
\begin{equation}
\label{genAij}
\hat A^{ij} = 
\hat A^{ij}_{{\bf C}_{1}{\bf P}_{1}} + \hat A^{ij}_{{\bf C}_{2}{\bf P}_{2}}.
\end{equation}
Equations~(\ref{BYAij}) or (\ref{genAij}) are now substituted back
into the Hamiltonian constraint Eq.~(\ref{hconsimple}), which is then
solved for the conformal factor.  This step requires {\em boundary
conditions}, which must enforce the existence of black holes in the
constructed initial data.  In this paper, we use the puncture method
\cite{BraB97} (see also \cite{BeiO94,BeiO96}) to accomplish this.  In
this approach, an appropriate (singular) piece is split off the
conformal factor analytically, and Eq.~(\ref{hconsimple}) is rewritten
as an elliptic equation for the remainder, which is continuous and
finite throughout.  Typically, this latter equation is solved
numerically.  We will instead construct an approximate but analytical
solution perturbatively, as described in the following Section.


\section{Perturbative Solutions}
\label{PERT}

We construct perturbative binary black hole solutions by separately
considering the effects of the boost and the companion on each black
hole, and then combining the results.  In Section \ref{pertboosted} we
first consider an isolated black hole of bare mass ${\mathcal M}$ with
boost ${\bf P}$ and construct a solution that is accurate to second order in
$\epsilon_P$, where
\begin{equation}
\epsilon_P \equiv \frac{P}{\mathcal M}.
\end{equation}
In Section \ref{pertcomp} we then consider a static black hole of bare
mass ${\mathcal M}_{1}$ with a companion of bare mass ${\mathcal
M}_{2}$ at a coordinate distance $s$ and find a solution that is
accurate to first order in $\epsilon_s$, where
\begin{equation}
\epsilon_s \equiv \frac{{\mathcal M}}{s}.
\end{equation}
In order to avoid unnecessarily complicated notation we do not
distinguish between ${\mathcal M}_1$ and ${\mathcal M}_2$ in
$\epsilon_P$ and $\epsilon_s$, and instead simply point out that
these expressions do depend on the mass ratio.

Finally, in Section \ref{pertbinary} we combine results to find
perturbative binary black hole solutions.  
For systems in equilibrium
the Virial theorem implies that we must have
\begin{equation}
\epsilon_s \approx \epsilon_P^2,
\end{equation}
so that  the orders of expansion for our treatment of the boost and the
companion are consistent with each other.


\subsection{A Single Boosted Black Hole}
\label{pertboosted}

Consider a static black hole with bare mass ${\mathcal M}$ at
coordinate location ${\bf C}$.  The asymptotically flat solution to
the Hamiltonian constraint (\ref{hconsimple}) is then the well-known
expression
\begin{equation}
\psi= 1 + \frac{{\mathcal M}}{2r_{{\bf C}}},
\end{equation}
describing a Schwarzschild black hole in isotropic coordinates.

To generalize this solution for a boosted black hole with momentum
${\bf P}$ we adopt the puncture method and split the conformal factor
into two terms,
\begin{equation} \label{con_factor_P}
\psi=\frac{1}{\alpha}+u_{P}.
\end{equation}
Here
\begin{equation}
\frac{1}{\alpha}=1+\frac{{\mathcal M}}{2r_{{\bf C}}}
\end{equation}
absorbs the singular term analytically, and $u_{P}$ is a regular
correction term that accounts for the effects of the boost.  In terms
of $u_{P}$ the Hamiltonian constraint (\ref{hconsimple}) becomes
\begin{equation}
\label{pertboosthamu}
\hat \nabla^{2}u_{P}=-\beta\left(1+\alpha u_{P}\right)^{-7},
\end{equation}
where
\begin{equation}
\label{pertboostedbeta}
\beta=\frac{1}{8}\alpha^{7}\hat A_{ij}^{{\bf CP}}\hat A^{ij}_{{\bf CP}}.
\end{equation}
Since $\hat A^{ij}_{{\bf CP}}$ scales with $P$, the leading order term
in (\ref{pertboosthamu}) scales with $\epsilon_P^2$ and all odd-order
terms in $\epsilon_P$ must vanish.  An analytical solution to order
$\epsilon_P^2$ is given by
\begin{eqnarray}
\label{pertboostusol}
u_{P} & = & \frac{{\mathcal M}\epsilon_{P}^{2}}
{8\left({\mathcal M}+2r_{{\bf C}}\right)^{5}}
\Big( u_{0}\left(r_{{\bf C}}\right)P_{0}(\cos\theta)+\nonumber\\ 
& & u_{2}\left(r_{{\bf C}}\right)P_{2}(\cos\theta) \Big) + 
{\mathcal O}(\epsilon_{P}^{4}),
\end{eqnarray}
where
\begin{equation}
P_{0}(\cos\theta)=1
\end{equation}
and
\begin{equation}
P_{2}(\cos\theta)=\frac{3}{2}\cos^{2}\theta-\frac{1}{2}
\end{equation}
are Legendre polynomials and where the radial functions 
$u_{0}\left(r_{{\bf C}}\right)$ and $u_{2}\left(r_{{\bf C}}\right)$
are
\begin{equation}
u_{0}\left(r_{{\bf C}}\right)={\mathcal M}^{4}+
10{\mathcal M}^{3}r_{{\bf C}}+40{\mathcal M}^{2}r_{{\bf C}}^{2}+
80{\mathcal M}r_{{\bf C}}^{3}+80r_{{\bf C}}^{4},
\end{equation}
and
\begin{eqnarray}
u_{2}\left(r_{{\bf C}}\right)&=&
\frac{{\mathcal M}}{5r^{3}_{\bf C}}
\Big( 42{\mathcal M}^{5}r_{{\bf C}}+378{\mathcal M}^{4}r_{{\bf C}}^{2}+
1316{\mathcal M}^{3}r_{{\bf C}}^{3}+ \nonumber \\ 
&& 2156{\mathcal M}^{2}r_{{\bf C}}^{4}+
1536{\mathcal M}r_{{\bf C}}^{5}+ 240r_{{\bf C}}^{6}+\nonumber \\
& &
21{\mathcal M}\left({\mathcal M}+2r_{\bf C}\right)^{5}
\ln\big(\frac{{\mathcal M}}{{\mathcal M}+2r_{{\bf C}}}\big) \Big)
\end{eqnarray}
(see Appendix \ref{Appboostedpsi} as well as \cite{Lag04}).  Expanding the last term of 
$u_{2}\left(r_{{\bf C}}\right)$ about $r_{{\bf C}}=0$ shows that $u_{2}(r_{{\bf C}})={\cal O}(r_{\bf C}^2)$.

>From the conformal factor (\ref{con_factor_P}) we can now compute
several quantities of interest.  A short calculation in appendix
\ref{Appboostedmadm} shows that the ADM energy $E$ of the solution is
\begin{equation}
\label{pertboostEofbare}
E={\mathcal M}+\frac{5}{8}{\mathcal M}\epsilon_{P}^{2}+
{\mathcal O}(\epsilon_{P}^{4})=
{\mathcal M}+\frac{5P^{2}}{8{\mathcal M}}+{\mathcal O}(\epsilon_{P}^{4}).
\end{equation}
We would also like to know the irreducible mass
\begin{equation}
M = \sqrt{\frac{A}{16 \pi}},
\end{equation}
where $A$ is the proper area of the black hole's event horizon
\cite{Chr70}.  Since we cannot determine the location of the event
horizon from initial data alone, we approximate this area by the area
of the apparent horizon.  One might expect that computing the
irreducible mass to second order in $\epsilon_P$ requires the position
of the horizon to second order as well.  As it turns out, however, the
second order term in the horizon's position cancels out (see Appendix
\ref{Appboostedmirr}), so that we only need the first order correction
(see Appendix \ref{Appboostedh})
\begin{equation}
h = \frac{{\mathcal M}}{2}-
\frac{{\mathcal M}}{16}\epsilon_{P}\cos\theta+{\mathcal O}(\epsilon_{P}^{2}).
\end{equation}
As we show in Appendix
\ref{Appboostedmirr}, the irreducible mass $M$ is then given by
\begin{equation}
\label{pertboostirrofbare}
M={\mathcal M}+\frac{1}{8}{\mathcal M}\epsilon_{P}^{2}+
{\mathcal O}(\epsilon_{P}^{4})=
{\mathcal M}+\frac{P^{2}}{8{\mathcal M}}+{\mathcal O}(\epsilon_{P}^{4}).
\end{equation}
We have verified the results for the ADM energy $E$
(\ref{pertboostEofbare}) and irreducible mass $M$
(\ref{pertboostirrofbare}) by comparing them with numerical results
obtained with a pseudo-spectral elliptic solver~\cite{PfeKST03}.
Spectral techniques are well suited for this purpose, since they
provide sufficient accuracy to resolve the higher-order terms
neglected in the analytic treatment.  We note that the spectral
elliptic solver appears to handle puncture data well, despite initial
concerns about low differentiablility of $u$ at the punctures.  This
observation would benefit from further investigation.

\begin{figure}
\includegraphics[width=0.48\textwidth]{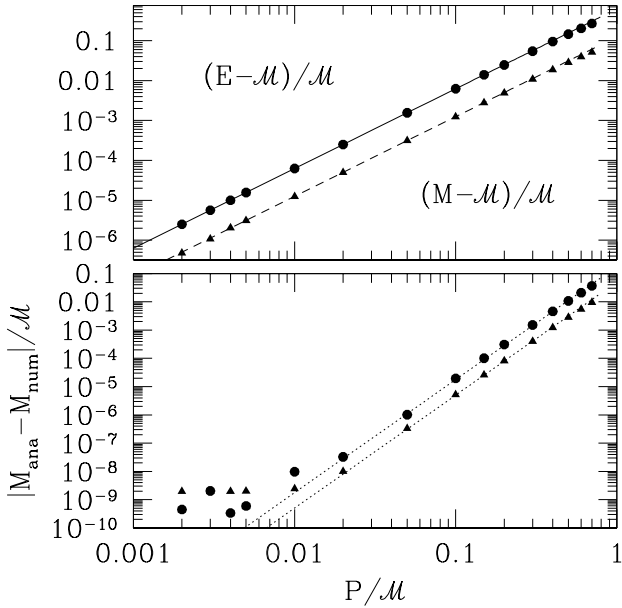}
\caption{The top panel shows the ADM energy (solid line) and
irreducible mass (dashed) for a boosted black hole.  The lines are
drawn according to equations (\ref{pertboostEofbare}) and
(\ref{pertboostirrofbare}); the points are numerical results, with
error bars smaller than the points.  The bottom panel shows the
difference between our analytical and numerical results.  The dotted
lines have slopes of $4$, indicating that the errors in our analytical
results are of order $\epsilon_P^4$.  The deviation from this
scaling for very small values of $P/{\mathcal M}$ is caused by the
truncation error in the numerical data.}
\label{nummasses}
\end{figure}

In Fig~\ref{nummasses} we show $E$ and $M$ as a function of $P$,
including both perturbative and numerical values and their difference.
Since the leading order error of the perturbative results scales with
$\epsilon_P^4$ we find remarkably good agreement even for moderately
large values of the momentum $P$.  For conformally flat Bowen-York
based initial data, $P/M \approx 0.6$ at the predicted location of the
innermost stable circular orbit (ISCO, e.g.~\cite{Coo94,Bau00}), at
which point the relative error of the perturbative values is
approximately two percent for the ADM energy, and less than one
percent for the irreducible mass.  There is general consensus that the
ISCO occurs at a somewhat larger binary separation
(e.g.~\cite{CauCGP06}), where both $P$ and our errors are even
smaller.

\begin{figure}
\includegraphics[width=0.48\textwidth]{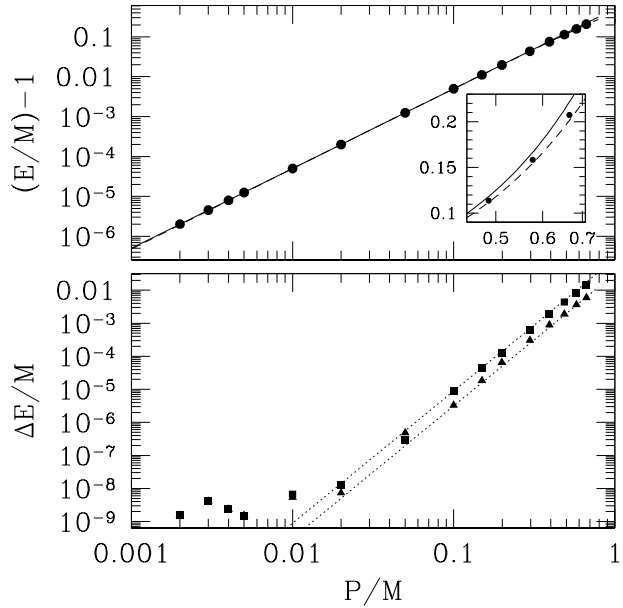}
\caption{The top panel shows perturbative (solid line,
Eq.~(\ref{newtenergy})) and numerical (dots) results for the ADM
energy of a boosted black hole.  The dashed line shows the ADM energy
of a properly boosted Schwarzschild black hole, Eq.~(\ref{ESch}).  The
bottom panel shows the deviation $(E_{\rm num}-E_{\rm Sch})/M$
introduced by representing the boosted black hole with conformally
flat, maximally sliced puncture data (triangles) as well as the 
error $|E_{\rm ana}-E_{\rm num}|/M$ introduced by our
perturbative solution (squares).  The dotted lines have slopes of $4$,
indicating that all errors scale as $\epsilon_P^4$.}
\label{gammaplot}
\end{figure}

In (\ref{pertboostEofbare}) the ADM energy $E$ is given in terms of
the bare mass ${\mathcal M}$, which has a physical meaning only in the
limit of infinite separation.  It is more intuitive to express $E$ in
terms of the black hole's irreducible mass.  Inverting
(\ref{pertboostirrofbare}) we have
\begin{equation}
{\mathcal M}=M-\frac{P^{2}}{8M}+{\mathcal O}(\epsilon_{P}^{4}),
\end{equation}
which we can now insert into (\ref{pertboostEofbare}) to find
\begin{equation}
\label{newtenergy}
E=M+\frac{P^{2}}{2M}+{\mathcal O}(\epsilon_{P}^{4}).
\end{equation}
This result is not surprising but reassuring.  As expected, we can
interpret the ADM energy as the sum of a ``rest mass'' -- identified with the
irreducible mass -- and a Newtonian kinetic energy term.  Given that
we only work to order $\epsilon_P^2$ we expect to find these Newtonian
expressions only.  It is worth emphasizing, however, that these
Newtonian expressions emerge only after carefully taking into account
the effect of the boost on the black hole's irreducible mass, which is
an intrinsically relativistic object.

In Fig.~\ref{gammaplot} we show perturbative and numerical results for
the ADM energy as function of $P/M$.  We also include the special
relativistic result for the ADM energy of a
properly boosted Schwarzschild black hole,
\begin{equation}\label{ESch}
E_{\rm Sch}=\sqrt{M^2+P^2}=M+\frac{P^2}{2M}
-\frac{P^4}{8M^3}+{\cal O}(\epsilon_P^6).
\end{equation}
Comparing the values for $E_{\rm ana}$, $E_{\rm num}$, and $E_{\rm
Sch}$ we see that the error in the ADM energy introduced by
constructing transverse-traceless initial data {\em perturbatively}
instead of exactly is only somewhat larger than the deviation of the
ADM energy of conformally flat transverse-traceless puncture data from
 $E_{\rm Sch}$.  Both deviations scale as $P^4$ and represent an
excess energy over the Schwarzschild value that can be interpreted as
being associated with gravitational radiation.


\subsection{A Static Black Hole with a Companion}
\label{pertcomp}

We would like to derive a result as familiar as (\ref{newtenergy}) for a 
static black hole with bare mass ${\mathcal M}_{1}$ in the presence of 
a second static black hole with bare mass ${\mathcal M}_{2}$ a coordinate
distance $s$ away.  For this system, the exact solution to the
Hamiltonian constraint (\ref{hconsimple}) is
\begin{equation}
\label{pertcomppsiexact}
\psi=1+\frac{{\mathcal M}_{1}}{2r_{{\bf C}_{1}}}+
\frac{{\mathcal M}_{2}}{2r_{{\bf C}_{2}}}.
\end{equation}
In equation (\ref{pertcomppsiexact}) $r_{{\bf C}_{1}}$ is the
coordinate distance from the center of hole ${\mathcal M}_{1}$, and
$r_{{\bf C}_{2}}$ is the coordinate distance from the center of hole
${\mathcal M}_{2}$.  Evaluating the ADM energy (\ref{ADM}) for the conformal
factor (\ref{pertcomppsiexact}) yields
\begin{equation} \label{pertcompEofbare}
E={\mathcal M}_{1}+{\mathcal M}_{2}.
\end{equation}

\begin{figure}
\includegraphics[width=0.48\textwidth]{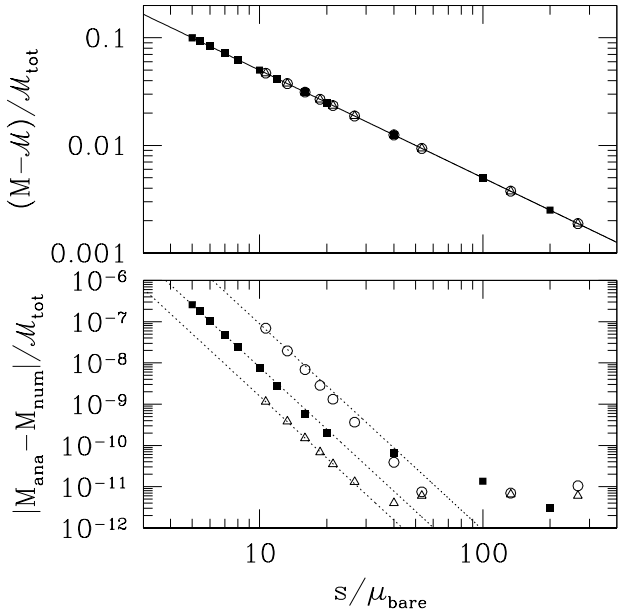}
\caption{The top panel shows the irreducible mass for a black hole
with a companion a coordinate distance $s$ away.  Two cases with the
same total bare mass ${\mathcal M}_{\rm tot}={\mathcal
M}_{1}+{\mathcal M}_{2}$ but different reduced bare masses $\mu_{\rm
bare}={\mathcal M}_{1}{\mathcal M}_{2}/{\mathcal M}_{\rm tot}$ are
shown: ${\mathcal M}_{1}={\mathcal M}_{2}=1$ and ${\mathcal
M}_{1}=\frac{3}{2}$, ${\mathcal M}_{2}=\frac{1}{2}$.  The line drawn
is the prediction of equation (\ref{pertcompmirr}), and is the same
for each case. The points are numerical results: squares for the equal
mass case, and circles and triangles for the larger and small masses,
respectively.  The bottom panel shows the difference between our
analytical and numerical results.  The dotted lines have slopes of
$-5$, indicating that the errors in our analytical results are of
order $\epsilon_s^5$.}
\label{figsep}
\end{figure}

To calculate the irreducible masses $M_{1}$ and $M_{2}$ to the order
$\epsilon_s$ necessary to reproduce the expected classical result for
$E$, we approximate $\psi$ in a neighborhood of ${\mathcal M}_{1}$
as
\begin{equation}
\label{comppsinear1}
\psi=1+\frac{{\mathcal M}_{1}}{2r_{{\bf C}_{1}}}+
\frac{{\mathcal M}_{2}}{2s}+{\mathcal O}(\epsilon_s^{2}),
\end{equation}
(see Appendix \ref{Appcomppsi}).  In a neighborhood of ${\mathcal
M}_{2}$, $\psi$ is given by the same expression with the indices
interchanged.  We would again like to evaluate the irreducible masses
$M_{1}$ and $M_{2}$ of the two black holes.  We proceed exactly as we
did for isolated boosted black holes, except that in this case we only
need to expand to order $\epsilon_s$.  Since the correction to the
location of the horizon $h$ only enters squared into the expression
for the irreducible mass (see (\ref{AppboostedA4terms}) in Appendix
\ref{Appboostedmirr}), we can completely neglect the perturbation of
the horizon and may take into account only the zeroth order term
\begin{equation}
h = \frac{{\mathcal M}_{1}}{2}+{\mathcal O}(\epsilon_{s}).
\end{equation}
We then find the irreducible mass
\begin{equation} \label{pertcompmirr}
M_{1}={\mathcal M}_{1} +
\frac{{\mathcal M}_{1}{\mathcal M}_{2}}{2s} +
{\mathcal O}(\epsilon_s^{2})
\end{equation}
(See Appendix \ref{Appcompmirr}).  Interchanging indices yields the
irreducible mass of black hole ${\mathcal M}_{2}$.  In
Figure~\ref{figsep} we show the irreducible masses of static black
holes with companions as a function of binary separation $s$,
including both perturbative and numerical values and their difference.
We again find remarkably good agreement to very small binary
separations.  As shown in \cite{SchKB06}, the expression
(\ref{pertcompmirr}) turns out to be accurate to at least third order
in $\epsilon_s$, and our comparison in Fig.~\ref{figsep} demonstrates
that the leading order error term scales with $\epsilon_s^5$.  As
computed from the transverse-traceless decomposition, for equal masses
the ISCO occurs at $s/M_{\rm tot} \approx 2.25$ (e.g.~\cite{Bau00}),
where the relative error is approximately $10^{-8}$.  This again sets
an upper limit, since more realistically the ISCO is believed to occur
at a somewhat larger binary separation.

As for the single boosted black holes we again invert
(\ref{pertcompmirr}) 
\begin{equation}
{\mathcal M}_{1}=M_{1}-\frac{M_{1}M_{2}}{2s}+
{\mathcal O}(\epsilon_s^{2}),
\end{equation}
and insert this expression into (\ref{pertcompEofbare}) to find the
ADM energy $E$ in terms of the irreducible mass
\begin{equation}
E=M_{1}+M_{2}-\frac{M_{1}M_{2}}{s}+{\mathcal O}(\epsilon_s^{2}).
\end{equation}
This is again the expected Newtonian result which allows us to interpret
the ADM energy as the sum of the ``rest mass'' of the two black holes, identified
with the irreducible masses, plus the potential energy.


\subsection{A Binary Black Hole System}
\label{pertbinary}

We are now in a position to construct perturbative binary black hole
solutions by combining the results of Sections \ref{pertboosted} and
\ref{pertcomp}.  As it turns out, we can simply add both results and
obtain a solution that is accurate to order 
\begin{equation} \label{order}
\epsilon \equiv \epsilon_s \approx \epsilon_P^2
\end{equation}
(where we assume that $\epsilon_s$ and $\epsilon_P$ are similar for
both black holes.)  More systematically, however, we proceed as
follows.

First, we would like to solve the Hamiltonian constraint
(\ref{hconsimple}) for two black holes: one at coordinate location
${\bf C}_{1}$ with bare mass ${\mathcal M}_{1}$ and momentum ${\bf
P}_{1}$, and the other at coordinate location ${\bf C}_{2}$ with bare
mass ${\mathcal M}_{2}$ and momentum ${\bf P}_{2}$.  As in Sections
\ref{pertboosted} and \ref{pertcomp} we start with the corresponding
static solution
\begin{equation}
\psi=1+\frac{{\mathcal M}_{1}}{2r_{{\bf C}_{1}}}+
\frac{{\mathcal M}_{2}}{2r_{{\bf C}_{2}}}
\end{equation}
and find a correction $u$ using the puncture method of \cite{BraB97}
for non-vanishing $\hat A^{ij}$.  Defining
\begin{equation}
\psi=u+\frac{1}{\alpha},
\end{equation}
with
\begin{equation} \label{pertcompalpha}
\frac{1}{\alpha}=1+\frac{{\mathcal M}_{1}}{2r_{{\bf C}_{1}}}+
\frac{{\mathcal M}_{2}}{2r_{{\bf C}_{2}}}
\end{equation}
the Hamiltonian constraint (\ref{hconsimple}) becomes a regular
equation for the correction $u$
\begin{equation}
\label{hconubinary}
\hat\nabla^{2}u=-\beta\left(1+\alpha u\right)^{-7}.
\end{equation}
Here we have again used
\begin{equation}
\beta=\frac{1}{8}\alpha^{7}\hat A_{ij}\hat A^{ij},
\end{equation}
and $\hat A_{ij}$ is now given by equation (\ref{genAij}).  Equation
(\ref{hconubinary}) is nonlinear, and is usually solved numerically.
We will solve it perturbatively, using our solutions from sections
\ref{pertboosted} and \ref{pertcomp}.  Intuition suggests, and
Appendix \ref{Appbinpsi} proves that to fourth order in the
momenta, $u$ is simply the sum of the boost corrections for each hole
taken separately,
\begin{equation}
u=u_{P_{1}}+u_{P_{2}}+{\mathcal O}(\epsilon^{2}),
\end{equation}
where $u_{P_{1}}$ and $u_{P_{2}}$ are the isolated black hole
perturbations given by equation (\ref{pertboostusol}).  The first
neglected terms are of order $\epsilon^{2}\approx \epsilon_{P}^{4}$
because, as in Section \ref{pertboosted}, $\hat A^{ij}$ scales with
$P$, so that the leading order term in (\ref{hconubinary}) scales with
$\epsilon_{P}^{2}$ and all odd-order terms in $\epsilon_{P}$ must
vanish.  The conformal factor is then
\begin{equation}
\label{pertbinpsisol}
\psi=1+\frac{{\mathcal M}_{1}}{2r_{{\bf C}_{1}}}+
\frac{{\mathcal M}_{2}}{2r_{{\bf C}_{2}}}+u_{P_{1}}+u_{P_{2}}+
{\mathcal O}(\epsilon^{2}),
\end{equation}
and it follows immediately that the ADM energy of the system to this order is
\begin{equation}
E={\mathcal M}_{1}+\frac{5P_{1}^{2}}{8{\mathcal M}_{1}}+
{\mathcal M}_{2}+\frac{5P_{2}^{2}}{8{\mathcal M}_{2}}+
{\mathcal O}(\epsilon^{2}).
\end{equation}

Finding the irreducible mass of each hole again requires finding the
apparent horizons surrounding each hole.  Since we only consider the
leading order perturbations of a Schwarzschild black hole, we can
simply add the contributions from the boost and the companion and find
\begin{equation}
h_{1} =\frac{{\mathcal M}_{1}}{2}-\frac{P_{1}}{16{\mathcal M}_{1}}\cos\theta_{1}
+{\mathcal O}(\epsilon),
\end{equation}
where $\theta_{1}$ is measured from ${\bf P}_{1}$.  We note in
particular that ${\mathcal M}_2$'s boost affects ${\mathcal M}_1$ only
at higher order.  The irreducible mass of the black hole ${\mathcal
M}_1$ is then
\begin{equation}
M_{1}={\mathcal M}_{1}+\frac{P_{1}^{2}}{8{\mathcal M}_{1}}+
\frac{{\mathcal M}_{1}{\mathcal M}_{2}}{2s}+
{\mathcal O}(\epsilon^2)
\end{equation}
(and similar for ${\mathcal M}_2$; see Appendix \ref{Appbinmirr}.)
The correction to the irreducible mass of each black hole is the sum
of the separate corrections for its boost and the presence (but not
the momentum) of its companion.  As before we can solve for the bare
masses,
\begin{equation}
\label{1bareirrbin}
{\mathcal M}_{1}=M_{1}-\frac{P_{1}^{2}}{8M_{1}}-
\frac{M_{1}M_{2}}{2s}+{\mathcal O}(\epsilon^2)
\end{equation}
and similar for ${\mathcal M}_2$,
and express the ADM energy $E$ in terms of the irreducible masses
\begin{equation}
\label{Ecompirrs}
E=M_{1}+M_{2}+\frac{P_{1}^{2}}{2M_{1}}+\frac{P_{2}^{2}}{2M_{2}}-
\frac{M_{1}M_{2}}{s}+ {\mathcal O}(\epsilon^2).
\end{equation}
Again, this result is not surprising but reassuring.


\subsection{Quasicircular Orbits}
\label{QCO}

Because of the circularizing effects of gravitational radiation,
binary black holes at reasonably small binary separation are expected
to be in approximately circular orbit.  We can construct such systems
by minimizing the binding energy while keeping the orbital angular
momentum and irreducible masses fixed (see, e.g., \cite{Bau01} for an
illustration.)  Since Eq.~(\ref{Ecompirrs}) represents the
Newtonian energy, we will recover expressions of Newtonian orbits in
what follows, but we also keep track of the error term.  We define the
binding energy as the difference between the ADM energy $E$ and the
black holes' irreducible masses,
\begin{equation}
\label{QCOEbdef}
E_b = E-M_{1}-M_{2}.
\end{equation}
We also define the total mass
\begin{equation}
M_{{\rm tot}}=M_{1}+M_{2}
\end{equation}
and the reduced mass
\begin{equation}
\mu=\frac{M_{1}M_{2}}{M_{1}+M_{2}}.
\end{equation}
In a reference frame where the total momentum is zero both individual
momenta ${\bf P}_1$ and ${\bf P}_2$ must have the same magnitude
\begin{equation}
P_{1}=P_{2} \equiv P.
\end{equation}
In such a frame we can then write the binding energy (\ref{QCOEbdef}) as
\begin{equation} \label{QCOEb}
E_{b}=\frac{P^{2}}{2\mu}-
\frac{M_{{\rm tot}}\mu}{s}+{\mathcal O}(\epsilon^{2}).
\end{equation}
Not surprisingly, this is again the Newtonian expression for the binding
energy, and as a consequence the entire following discussion is essentially
Newtonian.

Quasicircular orbits are those satisfying
\begin{equation}
\left. \frac{\partial E_{b}}{\partial s} \right|_{M_{1},M_{2},J}=0,
\end{equation}
where $J=Ps$ is the orbital angular momentum (note that $J={\cal
O}(\epsilon^{-1/2})$ ).  In terms of $J$,
\begin{equation}
\label{ebj}
E_{b}=\frac{J^{2}}{2s^{2}\mu}-
\frac{M_{{\rm tot}}\mu}{s}+{\mathcal O}(\epsilon^2).
\end{equation}
Taking the derivative, we find that quasicircular orbits are those for which
\begin{equation} \label{J}
\frac{J}{\mu M_{\rm tot}} = \left(\frac{M_{\rm tot}}{s}\right)^{-1/2} + {\mathcal O}(\epsilon^{1/2}),
\end{equation}
or
\begin{equation}
\label{qcp}
\frac{P}{\mu} = \left(\frac{M_{\rm tot}}{s}\right)^{1/2}+{\mathcal O}(\epsilon^{3/2}).
\end{equation}
This condition is equivalent to a Virial relation and justifies our
relation (\ref{order}).  Consequently, quasicircular orbits have a
binding energy
\begin{equation} \label{E_b}
\frac{E_{bQC}}{\mu}=-\frac{1}{2}\frac{M_{{\rm tot}}}{s}+{\mathcal O}(\epsilon^{2}),
\end{equation}
as one might have expected.  The orbital angular frequency, $\Omega$,
measured at infinity, is
\begin{equation}
\Omega= \left. \frac{\partial E_{b}}{\partial J}
\right|_{M_{1},M_{2},s}.
\end{equation}
Inserting the binding energy (\ref{ebj}) we find
\begin{equation}
\label{Kepler}
M_{\rm tot}\Omega = \left(\frac{M_{\rm tot}}{s}\right)^{3/2}+{\mathcal O}(\epsilon^{5/2}),
\end{equation}
which we recognize as Kepler's third law.  Substituting into
Eq.~(\ref{E_b}), we finally find
\begin{equation}\label{E_b-vs-Omega}
\frac{E_{bQC}}{\mu}=-\frac{1}{2}\left(M_{\rm tot}\Omega\right)^{2/3}
+{\cal O}\left( (M_{\rm tot}\Omega)^{4/3}\right).
\end{equation}

\begin{figure}
\includegraphics[width=0.48\textwidth]{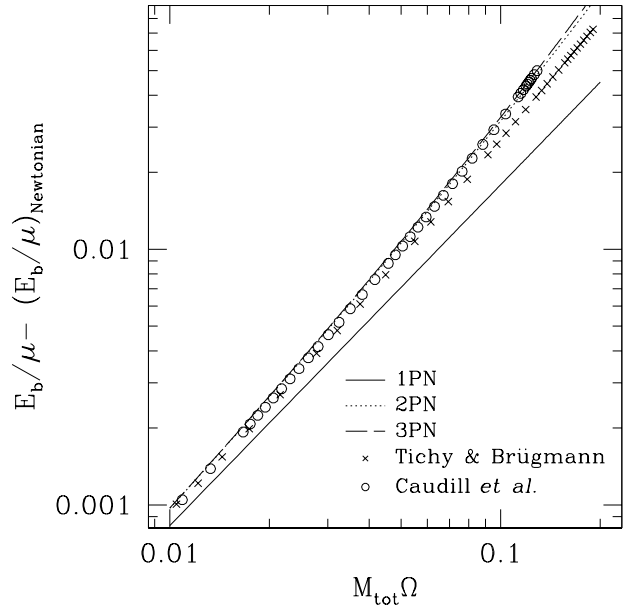}
\caption{The deviation in binding energy $E_b/\mu$ of an equal mass
binary between our calculation (which yields the Newtonian result) 
and more sophisticated approaches as a
function of orbital frequency.  We compare with the
post-Newtonian expansion of~\cite{Bla06} as well as numerical relativity
results~\cite{TicB04,CauCGP06}.  The line for ``1PN'' has precisely the
slope of our error term, $(M_{\rm tot}\Omega)^{4/3}$, confirming that
our analysis finds the correct scaling. }
\label{figbin}
\end{figure}

Evidently, our expressions for the angular momentum, the binding
energy and the orbital angular frequency are, as expected, simply the
Newtonian point-mass expressions.  Consequently, our perturbative
approach gives asymptotically correct results at sufficiently large
separations.  The errors in our approach are dominated by the first
post-Newtonian contributions.  As an illustration, Fig.~\ref{figbin}
shows the difference between our result for the binding energy of the
binary, Eq.~(\ref{E_b-vs-Omega}), and the results of more elaborate
schemes, namely post-Newtonian expansions~\cite{Bla06} and numerically
constructed BBH sequences~\cite{TicB04,CauCGP06}.  The errors in our
expressions scale precisely as expected as $(M_{\rm
tot}\Omega)^{4/3}\sim (M_{\rm tot}/s)^2$.

The graph demonstrates that for a given binary separation the
deviations in $E_b/\mu$ are larger than the deviations for the
individual contributions of the boost and the companion,
cf.~Figs.~\ref{gammaplot} and~\ref{figsep}.  This indicates that those
${\cal O}(\epsilon^2)$ terms which we neglected when constructing the
binary (cf.~Sec.~\ref{pertbinary}) have larger coefficients than those
for single boosted or tidally deformed black holes.



\section{Constructing Initial Data}
\label{cookbook}

In this Section we briefly describe how, for arbitrary irreducible
masses $M_1$ and $M_2$ and binary separation $s$, we can use the
results of the previous Sections to construct approximate binary black
hole initial data that are accurate to order $\epsilon$.

Without loss of generality we assume the orbital plane to be the
$z=0$ plane of a Cartesian coordinate system, and assume the $x$-axis
to connect the two holes.  For circular orbits the momenta ${\bf
P}_{1,2}$ must then be aligned with the $y$-axis.  In a reference
frame where the total momentum is zero we must also have
\begin{equation}
{\bf P}_{1}=-{\bf P}_{2}.
\end{equation}
>From (\ref{qcp}) the magnitude of the momenta is given by
\begin{equation}
\label{constr1}
P_{1}=P_{2}=\frac{\mu}{\sqrt{s/M_{\rm tot}}}
+{\mathcal O}(\epsilon^{2}).
\end{equation}
Using (\ref{1bareirrbin}) we find the bare mass
\begin{equation}
\label{constr2}
{\mathcal M}_{1}=M_{1}\left(1-\frac{M_{2}^{2}}{8s
\left(M_{1}+M_{2}\right)}-\frac{M_{2}}{2s}\right)+
{\mathcal O}(\epsilon^{2})
\end{equation}
and ${\mathcal M}_{2}$ is given by the same expression with the
indices interchanged.  To place the center of (irreducible) mass at
the origin of a Cartesian coordinate system, we choose 
\begin{equation}
\label{constr4}
{\bf C}_{1} = \left(\frac{M_{2}s}{M_{1}+M_{2}},0,0\right)
\end{equation}
for ${\mathcal M}_1$ and
\begin{equation}
\label{constr5}
{\bf C}_{2} = \left(-\frac{M_{1}s}{M_{1}+M_{2}},0,0\right)
\end{equation}
for ${\mathcal M}_2$.  With the momenta and positions of the two black
holes given, the extrinsic curvature $\hat A_{ij}$ can now be computed
from (\ref{genAij}) and (\ref{BYAij}).  Finally, the conformal factor
$\psi$ is given (\ref{pertbinpsisol}) and (\ref{pertboostusol}).  This
completes the construction of our approximate binary black hole initial
data.


\section{Summary}
\label{SUM}

We construct approximate but analytical binary black hole initial
data.  Adopting the puncture method (see \cite{BraB97,BeiO94,BeiO96})
we solve the constraint equations of general relativity in the
transverse-traceless decomposition for perturbations of isolated,
static Schwarzschild black holes, caused by a boost and the presence
of a binary companion.  We then use a superposition of the individual
perturbations to construct analytical black hole binary solutions that
are accurate to first order in the square of the momenta and the
inverse of the binary separation.  In particular, our initial data
become exact in the limit of infinite separation.

For binary black hole solutions constructed from the
transverse-traceless decomposition of the constraint equations,
quasi-circular orbits are usually identified by locating a minimum of
the binding energy along sequences of constant mass and angular
momentum (see Section \ref{QCO}).  For numerical solutions this
requires a large number of iterations, making the problem quite
involved.  In contrast, for our analytical solutions this becomes very
simple.  In Section \ref{cookbook} we describe how approximate binary
black hole solutions in quasicircular orbit, for arbitrary mass ratio,
can be set up analytically.  Given the amount of work required to
construct numerical solutions, our approximate solutions might be an
attractive alternative for some applications, especially for large
binary separations where our approximation becomes increasingly
accurate.   


Given that we expand only to first order in the square of the momentum
and the inverse of the binary separation, all expressions for global
quantities like the energy, angular momentum and orbital frequency
take their Newtonian values.  This does not mean, however, that our
calculation is Newtonian.  In fact, the above quantities only take the
expected Newtonian values after carefully taking into account
corrections to the black holes' irreducible mass, which is an
intrinsically relativistic quantity.  This study was partly motivated
by our curiousity in the behavior of the bare mass along sequences of
constant irreducible mass (compare footnote [35] and the surrounding
discussion in \cite{BauSS04}.)  Moreover, our solutions provide
approximate values for the conformal factor and the extrinsic
curvature together with the above global quantities, thereby
specifying complete sets of initial data describing binary black holes
in approximate quasicircular orbit.

\acknowledgments

TWB thanks the California Institute of Technology for hospitality
while this paper was finished.  HPP gratefully acknowledges support
through a Sherman Fairchild Prize fellowship.  The elliptic solver was
developed jointly with Lawrence Kidder and Mark Scheel.  This paper
was supported in part by NSF Grant No.~PHY-0456917 to Bowdoin College
and NSF grant PHY-0244906 to the California Institute of Technology.


\begin{appendix}


\section{A Boosted Black Hole}
\label{Appboosted}

\subsection{The Conformal Factor}
\label{Appboostedpsi}

We would like to solve the Hamiltonian constraint (\ref{hconsimple})
for a boosted black hole with bare mass ${\mathcal M}$ and momentum
${\bf P}$ located at ${\bf C}$.  We follow the puncture approach of
\cite{BraB97} and write the conformal factor $\psi$ as
\begin{equation}
\psi = \frac{1}{\alpha} + u_P
\end{equation}
(see Section \ref{pertboosted}), in which case the problem reduces
to solving
\begin{equation}
\label{Appboostedhamu}
\hat\nabla^{2}u_{P}=-\beta(1+\alpha u_{P})^{-7},
\end{equation}
for the correction term $u_{P}$.  Here the coefficients $\alpha$ and
$\beta$ are
\begin{equation}
\alpha= \left( 1+\frac{{\mathcal M}}{2r_{{\bf C}}} \right)^{-1},
\end{equation}
and
\begin{equation} \label{beta}
\beta=\frac{1}{8}\alpha^{7}\hat A_{ij}^{{\bf CP}}\hat A^{ij}_{{\bf CP}}.
\end{equation}
The extrinsic curvature $\hat A_{ij}^{{\bf CP}}$ is given by (\ref{BYAij}) and
yields
\begin{equation}
\hat A_{ij}^{{\bf CP}}\hat A^{ij}_{{\bf CP}}=
\frac{9 P^2}{r_{{\bf C}}^{4}} \left( \frac{1}{2}+\cos^{2}\theta \right),
\end{equation}
where $\theta$ is measured from ${\bf P}$.  Since $P$ enters into the
equations only squared, all odd order terms in $P$ must vanish.  We
construct a perturbative solution that is accurate to order $P^2$,
so that the leading order error term scales with $P^4$.  Up to this
order, (\ref{Appboostedhamu}) reduces to
\begin{equation}
\hat\nabla^{2}u_{P}+\beta+{\mathcal O}(\epsilon_{P}^{4})=0
\end{equation}
where $\epsilon_P = P/{\mathcal M}$.  Because of the axisymmetry of
this problem, it is reasonable to guess a solution of the form
\begin{equation} \label{up}
u_{P} =\epsilon_{P}^{2} \left( 
\tilde u_{0}(r_{{\bf C}})P_{0}(\cos\theta)+
\tilde u_{2}(r_{{\bf C}})P_{2}(\cos\theta) \right)+
{\mathcal O}(\epsilon_{P}^{4}),
\end{equation}
where $P_{0}(\cos\theta)=1$ and $P_{2}(\cos\theta)=
(3\cos^{2}\theta-1)/2$ are Legendre polynomials.
Solutions satisfying the boundary conditions $u_{P}\rightarrow0$ as
$r_{{\bf C}}\rightarrow\infty$ and $(\partial u_{P}/\partial
r_{{\bf C}})|_{r_{{\bf C}}=0}=0$ are
\begin{eqnarray}
\tilde u_{0}\left(r_{{\bf C}}\right)&=&\frac{{\mathcal M}}{8({\mathcal
M}+2r_{{\bf C}})^{5}} \Big( {\mathcal M}^{4}+10{\mathcal M}^{3}r_{{\bf
C}}+40{\mathcal M}^{2}r_{{\bf C}}^{2}\nonumber \\ & & +80{\mathcal
M}r_{{\bf C}}^{3}+80r_{{\bf C}}^{4} \Big),
\end{eqnarray}
and
\begin{eqnarray}
\tilde u_{2}\left(r_{{\bf C}}\right)&=& \frac{{\mathcal
M}^{2}}{40r_{{\bf C}}^{3}({\mathcal M}+2r_{{\bf C}})^{5}} \Big( 42{\mathcal
M}^{5}r_{{\bf C}}+378{\mathcal M}^{4}r_{{\bf C}}^{2}+\nonumber\\ & &
1316{\mathcal M}^{3}r_{{\bf C}}^{3}+2156{\mathcal M}^{2}r_{{\bf
C}}^{4}+1536{\mathcal M}r_{{\bf C}}^{5}+\nonumber\\ & & 240r_{{\bf
C}}^{6}+\nonumber \\ & & 21{\mathcal M}\left({\mathcal M}+2r_{\bf
C}\right)^{5}\ln\big(\frac{{\mathcal M}}{{\mathcal M}+2r_{{\bf
C}}}\big) \Big)
\end{eqnarray}
(see also \cite{Lag04}).  In Section \ref{pertboosted} we have
factored out terms that are common to both $\tilde u_{0}(r_{{\bf C}})$
and $\tilde u_{2}(r_{{\bf C}})$ (see (\ref{pertboostusol}).)


\subsection{The Apparent Horizon}
\label{Appboostedh}

In axisymmetry, the apparent horizon is located at a distance $r_{{\bf
C}}(\theta,\phi)=h(\theta)$ from a point ${\bf C}$ located on the axis of
symmetry, where $h(\theta)$ satisfies the following ordinary
differential equation \cite{Epp77}
\begin{eqnarray}
\label{horizonequ}
\partial_{\theta}\partial_{\theta} h & = & 
-\Gamma^{A}_{BC}m_{A}u^{B}u^{C}-
(\frac{ds}{d\theta})^{2}\gamma^{\phi\phi}\Gamma^{A}_{\phi\phi}m_{A}\nonumber \\
& & -(\gamma^{(2)})^{-1/2}(\frac{ds}{d\theta})u^{A}u^{B}K_{AB}-\nonumber \\ & &
(\gamma^{(2)})^{-1/2}(\frac{ds}{d\theta})^{3}\gamma^{\phi\phi}K_{\phi\phi},
\end{eqnarray}
subject to the boundary condition
\begin{equation}
\frac{\partial h}{\partial \theta}|_{\theta=0,\pi}=0.
\end{equation}
In (\ref{horizonequ}) we use the vectors
\begin{equation}
m_{i}=(1,-\partial_{\theta}h,0),
\end{equation}
and
\begin{equation}
u^{i}=(\partial_{\theta}h,1,0).
\end{equation}
We also abbreviate
\begin{equation}
(\frac{ds}{d\theta})^{2}=\gamma_{AB}u^{A}u^{B},
\end{equation}
where capital letters $A$, $B$, etc. run over the coordinates $r_{{\bf
C}}$ and $\theta$, but not $\phi$.  All coefficients are evaluated at
the horizon location, $r_{{\bf C}}=h$.  Given our assumption of
conformal flatness we also have
\begin{equation}
\gamma_{ij}=\psi^{4}\left( \begin{array}{ccc}
1 & 0 & 0 \\
0 & r_{{\bf C}}^{2} & 0 \\
0 & 0 & r_{{\bf C}}^{2}\sin^{2}\theta \end{array} \right),
\end{equation}
The black hole's momentum enters (\ref{horizonequ}) through the
extrinsic curvature $K_{ij}$.

It is easy to verify that for vanishing momentum $P = 0$ the well-known
isotropic horizon radius  
\begin{equation}
h_0 = \frac{\mathcal M}{2} 
\end{equation}
satisfies (\ref{horizonequ}) and the boundary conditions.  To find
a perturbative solution for non-zero $P$ we expand
\begin{equation} \label{h_expand}
h=h_{0}+h_{P}+h_{P^{2}}+{\mathcal O}(\epsilon_{P}^{3}),
\end{equation}
where $h_{P}$ and $h_{P^{2}}$ are corrections of first and second
order in $\epsilon_{P}$, respectively.

To find the first order correction $h_P$ we need to express all terms
in  (\ref{horizonequ}) up to order $\epsilon_P$, which yields
\begin{eqnarray}
\partial_{\theta}\partial_{\theta}h & = & 
\partial_{\theta}\partial_{\theta}h_{P}+{\mathcal O}(\epsilon_{P}^{2}), 
\nonumber \\
\Gamma^{A}_{BC}m_{A}u^{B}u^{C} & = & 
-\frac{h_{P}}{2}+{\mathcal O}(\epsilon_{P}^{2}), \nonumber \\
(\frac{ds}{d\theta})^{2}\gamma^{\phi\phi}\Gamma^{A}_{\phi\phi}m_{A} & = & 
(\cot\theta) \partial_{\theta}h_{P}-\frac{h_{P}}{2}+
{\mathcal O}(\epsilon_{P}^{2}), \nonumber \\
(\gamma^{(2)})^{-1/2}(\frac{ds}{d\theta})u^{A}u^{B}K_{AB} & = & 
\psi^{-2}K_{\theta\theta}+{\mathcal O}(\epsilon_{P}^{2}) \\ 
&=&-\frac{3P}{32}\cos\theta+{\mathcal O}(\epsilon_{P}^{2}),\nonumber \\
(\gamma^{(2)})^{-1/2}(\frac{ds}{d\theta})^{3}\gamma^{\phi\phi}K_{\phi\phi} 
& = &\psi^{-2}\sin^{-2}\theta K_{\phi\phi}+{\mathcal O}(\epsilon_{P}^{2}) 
\nonumber \\
&=&-\frac{3P}{32}\cos\theta+{\mathcal O}(\epsilon_{P}^{2}). \nonumber
\end{eqnarray}
The resulting equation for $h_{P}$ is
\begin{equation}
\partial_{\theta}\partial_{\theta}h_{P}-h_{P}+
(\cot\theta)\partial_{\theta}h_{P}-\frac{3P}{16}\cos\theta+
{\mathcal O}(\epsilon_{P}^{2})=0.
\end{equation}
The solution satisfying the boundary conditions is 
\begin{equation} \label{hp}
h_{P} = -\frac{P}{16}\cos\theta+{\mathcal O}(\epsilon_{P}^{2})
\end{equation}
(compare \cite{CooY90}).  One might
expect that in order to find the irreducible mass to order
$\epsilon_P^2$ we also need the next order term $h_{P^2}$.  However,
as we will see in Section \ref{Appboostedmirr}, this second order
terms cancels out, so that in fact we only need the first order
correction $h_P$.


\subsection{The Irreducible Mass}
\label{Appboostedmirr}

We would now like to find the irreducible mass of a boosted black hole, to
second order in $\epsilon_{P}$.  We approximate the irreducible mass as
\begin{equation}
M = \sqrt{\frac{A}{16\pi}},
\end{equation}
where $A$ is the proper area of the apparent horizon of the black hole
\cite{Chr70}.  It can be shown (see, e.g., Appendix D in
\cite{BauCSST96}) that
\begin{equation} \label{AppboostedA}
A=\int_{0}^{2\pi}\int_{0}^{\pi}\psi^{4}r_{{\bf C}}^{2}
\left( 1+\frac{1}{r_{{\bf C}}^{2}}
\left(\frac{\partial h}{\partial \theta}\right)^{2}\right)^{1/2}
\sin\theta d\theta d\phi,
\end{equation}
where $\psi$ and $r_{{\bf C}}$ are evaluated at the horizon location, $r_{{\bf
C}}=h$.  From the expansion (\ref{h_expand}) of $h$ we have
\begin{equation}
\label{Appboosteddhdtheta}
\left(\frac{\partial h}{\partial \theta}\right)^{2}=
\left(\frac{\partial h_{P}}{\partial \theta}\right)^{2}
+{\mathcal O}(\epsilon_{P}^{3}),
\end{equation}
since $h_0$ is spherically symmetric.  The horizon area (\ref{AppboostedA})
therefore splits into the two terms 
\begin{eqnarray}\label{AppboostedA2}
A & = &\int_{0}^{2\pi}\int_{0}^{\pi}\psi^{4}r_{{\bf C}}^{2}
\sin\theta d\theta d\phi+ \\ 
& & \frac{1}{2}\int_{0}^{2\pi}\int_{0}^{\pi}\psi^{4}
\left(\frac{\partial h_{P}}{\partial\theta}\right)^{2}
\sin\theta d\theta d\phi+ {\mathcal O}(\epsilon_{P}^{3}).
\nonumber
\end{eqnarray}
We now insert the expansions
\begin{equation}
\psi=1+\frac{{\mathcal M}}{2r_{{\bf C}}}+u_{P}+{\mathcal O}(\epsilon_{P}^{4})
\end{equation}
and
\begin{equation}
r_{{\bf C}}=h=\frac{{\mathcal M}}{2}+h_{P}+h_{P^{2}}+
{\mathcal O}(\epsilon_{P}^{3}).
\end{equation}
Taking these terms to their respective powers we find
\begin{equation}
\psi^{4}=16\left(1-\frac{4h_{P}}{{\mathcal M}}-
\frac{4h_{P^{2}}}{{\mathcal M}}+\frac{14h_{P}^{2}}{{\mathcal M}^{2}}+
2u_{P}\right)+{\mathcal O}(\epsilon_{P}^{3}),
\end{equation}
and
\begin{equation}
r_{{\bf C}}^{2}=\frac{{\mathcal M}^{2}}{4}+
{\mathcal M}h_{P}+{\mathcal M}h_{P^{2}}+h_{P}^{2}+
{\mathcal O}(\epsilon_{P}^{3}),
\end{equation}
which we can now insert into (\ref{AppboostedA2}) to find
\begin{eqnarray}
\label{AppboostedA4terms}
A& = &16\pi {\mathcal M}^{2}+
16\pi {\mathcal M}^{2}\int_{0}^{\pi}u_{P}
\sin\theta d\theta \nonumber \\ 
& & 
+ 16\pi \int_{0}^{\pi} h_{P}^{2}
\sin\theta d\theta 
+ \pi \int_{0}^{\pi}\psi^{4}
\left(\frac{\partial h_{P}}{\partial\theta}\right)^{2}
\sin\theta d\theta \nonumber \\
& & + {\mathcal O}(\epsilon_{P}^{3}). 
\end{eqnarray}
Quite remarkably, the second order term $h_{P^{2}}$ has canceled out
of this expression.  

To evaluate the horizon area we insert $u_p$ (equation (\ref{up}))
from Appendix \ref{Appboostedpsi} and $h_P$ (equation (\ref{hp})) from
Appendix \ref{Appboostedh}.  Since we are only working to second order
in $\epsilon_P$ it is sufficient to evaluate $u_P$ at the unperturbed
horizon location $h_0$,
\begin{eqnarray}
u_{P}&=&\frac{P^{2}}{2560{\mathcal M}^{2}}\big( -3422+672\ln 2 +\nonumber\\
&&(11196-2016\ln 2)\cos^{2}\theta \big)+{\mathcal O}(\epsilon_{P}^{3}).
\end{eqnarray}
In yet another remarkable happenstance the unattractive log terms 
disappear when the integration is carried out, and we are left with
\begin{equation}
A=16\pi {\mathcal M}^{2}(1+\frac{P^{2}}{4{\mathcal M}^{2}})+
{\mathcal O}(\epsilon_{P}^{4}),
\end{equation}
and therefore
\begin{equation}
M={\mathcal M}(1+\frac{P^{2}}{8{\mathcal M}^{2}})+
{\mathcal O}(\epsilon_{P}^{4}).
\end{equation}
Even though we have carried out this calculation only to order
$\epsilon_P^2$, it is clear that neither the area nor the mass can
depend on the sign of $P$, so that all odd order terms must disappear.
We have also verified this by comparing with numerical data in
Fig.~\ref{nummasses}.


\subsection{The ADM Energy}
\label{Appboostedmadm}

The ADM energy $E$ is defined as
\begin{equation} \label{ADM}
E = -\frac{1}{2\pi}\oint_{\infty}\hat\nabla^{i}\psi d^{2}S_{i}.
\end{equation}
According to (\ref{con_factor_P}) we write the conformal factor 
as $\psi=1+{\mathcal M}/2r_{{\bf
C}}+u_{P}$.  The correction term $u_P$ is given by  (\ref{up}),
and evaluating its leading order term as $r_{{\bf C}}\rightarrow\infty$
we find
\begin{equation}
u_{P}\sim\frac{5P^{2}}{16{\mathcal M}r_{{\bf C}}}+{\mathcal O}(\epsilon_{P}^{4}).
\end{equation}
The ADM energy therefore becomes
\begin{equation}
E ={\mathcal M}+\frac{5P^{2}}{8{\mathcal M}}+{\mathcal O}(\epsilon_{P}^{4}).
\end{equation}


\section{A Static Black Hole with a Companion}
\label{Appcomp}

\subsection{The Conformal Factor}
\label{Appcomppsi}

The {\em exact} solution for the conformal factor $\psi$ describing two
static black holes is given by
\begin{equation} \label{psi_exact}
\psi = 1 + \frac{{\mathcal M}_{1}}{2r_{{\bf C}_{1}}} 
+ \frac{{\mathcal M}_{2}}{2r_{{\bf C}_{2}}}.
\end{equation}
Focusing on ${\mathcal M}_1$ we would like to eliminate the dependence
on $r_{{\bf C}_2}$, which we can do in terms of the expansion
\begin{equation}
\label{generatelp}
\frac{1}{r_{{\bf C}_{2}}}=
\frac{1}{s}\sum^{\infty}_{n=0}
\left(\frac{r_{{\bf C}_{1}}}{s}\right)^{n}P_{n}\left(\cos\theta\right).
\end{equation}
Here $\theta$ is the angle between ${\bf r_{C}}_{1}$ and ${\bf
r}_{12}={\bf C}_{2}-{\bf C}_{1}$.  Since we are only interested in
terms up to order $\epsilon_s = {\mathcal M}/s$ it is
sufficient to keep the $n=0$ term as long as we restrict analysis
to a neighborhood of ${\mathcal M}_1$, where $r_{{\bf C}_1} \approx
{\mathcal M}_1$.  Near 
${\mathcal M}_1$ we then have
\begin{equation}
\frac{1}{r_{{\bf C}_{2}}}=\frac{1}{s}+{\mathcal O}(\epsilon_s^{2}),
\end{equation}
and hence
\begin{equation}
\psi=1+\frac{{\mathcal M}_{1}}{2r_{{\bf C}_{1}}}+
\frac{{\mathcal M}_{2}}{2s}+{\mathcal O}(\epsilon_s^{2}),
\end{equation}
and similar for ${\mathcal M}_2$.  Borrowing the notation for boosted
black holes we denote
\begin{equation} \label{u_C}
u_{C_1} = \frac{{\mathcal M}_{2}}{2s} + {\mathcal O}(\epsilon_s^{2}),
\end{equation}
and similar for ${\mathcal M}_2$.


\subsection{The Irreducible Mass}
\label{Appcompmirr}

Evidently the effect of a companion on an otherwise spherical black
hole must be axisymmetric in nature.  To find the irreducible mass of
the perturbed black hole ${\mathcal M}_1$ we can therefore again
evaluate the integral (\ref{AppboostedA4terms}), with $u_P$ replaced
by $u_{C_1}$ (equation (\ref{u_C}).)  As before we can expand the
horizon location as
\begin{equation}
h = h_0 + h_s + h_{s^2} + {\mathcal O}(\epsilon_s^{3}).
\end{equation}
Unlike in the case of a boosted black hole, however, we are only
interested in terms up to order $\epsilon_s$.  Since the leading order
term $h_s$ scales with $\epsilon_s$ and enters into the irreducible
mass only squared, we may neglect its contributions.  The only terms
that remain in (\ref{AppboostedA4terms}) are therefore the first two.
Inserting (\ref{u_C}) we immediately find
\begin{equation}
A_{1}=16\pi{\mathcal M}_{1}^{2}(1+\frac{{\mathcal M}_{2}}{s})+
{\mathcal O}(\epsilon_s^{2}),
\end{equation}
and hence
\begin{equation}
M_{1}={\mathcal M}_{1}+\frac{{\mathcal M}_{1}{\mathcal M}_{2}}{2s}+
{\mathcal O}(\epsilon_s^{2}).
\end{equation}
Interchanging indices yields ${\mathcal M}_{2}$.





\section{A Binary Black Hole System}
\label{Appbin}

\subsection{Conformal factor}
\label{Appbinpsi}

We would now like to solve the equation 
\begin{equation} \label{Appbinhamu}
\hat\nabla^{2}u=-\beta+ {\mathcal O}(\epsilon^2)
\end{equation}
for two boosted black holes (here $\epsilon$ is defined in (\ref{order})).  
The source term $\beta$ is 
\begin{equation} \label{Appbinbeta}
\beta=\frac{1}{8}\alpha^{7}\hat A_{ij} \hat A^{ij},
\end{equation}
where $\alpha$ is defined in (\ref{pertcompalpha}) and the extrinsic
curvature $\hat A_{ij}$ is given by (\ref{genAij}) as the sum of
the individual terms for the two black holes.  Given that this source
term is non-linear, it is not immediately evident that we can simply
superpose two single boosted black hole solutions
\begin{equation} \label{uguess}
u=u_{P_{1}}+u_{P_{2}}+{\mathcal O}(\epsilon^2),
\end{equation}
each of which satisfies (\ref{Appbinhamu}) with their individual
source term, e.g.
\begin{equation}
\hat\nabla^{2} u_{P_{1}} = -\beta_1 + {\mathcal O}(\epsilon^2).
\end{equation}
The superposition (\ref{uguess}) only satisfies (\ref{Appbinhamu}) to
the required order if
\begin{equation}
\beta-\beta_{1}-\beta_{2}= {\mathcal O}(\epsilon^2).
\end{equation}
This, however, is indeed the case.  Any point in space is at least a
coordinate distance $s/2$ separated from one of the black holes.  For
concreteness, consider a point that is at least a distance $s/2$ 
separated from ${\mathcal M}_2$.  Expanding $\beta$ at this point 
shows
\begin{equation}
\beta = \beta_1 +  {\mathcal O}(\epsilon^2),
\end{equation}
and we also have $\beta_2 = {\mathcal O}(\epsilon^5)$.  Therefore, the
superposition (\ref{uguess}) is a solution to (\ref{Appbinhamu}) to
the required order.  

We point out that the non-linearity of $\beta$ prevents us from also
including spin in our analysis.  Allowing for non-zero spin amounts to
including additional terms in the extrinsic curvature, which then lead
to cross-terms in the source term (\ref{Appbinbeta}).  Perturbative
solutions describing stationary, spinning black holes exist, but
because of these cross-terms they cannot simply be added to the
perturbative solutions describing non-spinning, boosted black holes to
construct spinning, boosted black holes.
 
\subsection{Irreducible mass}
\label{Appbinmirr}

We finally argue why the correction to the irreducible mass of a boosted
black hole with a companion is the sum of the corrections for a boost 
and the presence of a companion.  In the neighborhood of 
${\mathcal M}_1$, for example, the conformal factor is given by
(\ref{pertbinpsisol}),
\begin{equation}
\psi=1+\frac{{\mathcal M}_{1}}{2r_{{\bf C}_{1}}}+
u_{C_1} + u_{P_1},
\end{equation}
where $u_{C_1} = {\mathcal M}_2/(2s) + {\mathcal O}(\epsilon^2)$, and
${\mathcal M}_1$'s horizon is located at
\begin{equation}
r_{{\bf C}_{1}}=h_{1}=\frac{{\mathcal M}_{1}}{2}+
h_{P}+ {\mathcal O}(\epsilon).
\end{equation}
As before, higher order corrections to the location of the horizon
drop out.  We can now proceed exactly as in Appendix
\ref{Appboostedmirr}, using the combined correction $v = u_{C_1} +
u_{P_1}$ to the conformal factor, and find
\begin{equation}
A_{1}=16\pi{\mathcal M}_{1}^{2}(1+\frac{P_{1}^{2}}{4{\mathcal M}_{1}^{2}}+
\frac{{\mathcal M}_{2}}{s})+{\mathcal O}(\epsilon^2).
\end{equation}
The irreducible mass is therefore
\begin{equation}
M_{1}={\mathcal M}_{1}(1+\frac{P_{1}^{2}}{8{\mathcal M}_{1}^{2}}+
\frac{{\mathcal M}_{2}}{2s})+{\mathcal O}(\epsilon^2)
\end{equation}
as expected.

\end{appendix}


\end{document}